\newcommand{\beq}{\begin{equation}}
\newcommand{\eeq}{\end{equation}}
\newcommand{\bea}{\begin{eqnarray}}
\newcommand{\eea}{\end{eqnarray}}

\newcommand{\gsim}{\lower.7ex\hbox{$\;\stackrel{\textstyle>}{\sim}\;$}}
\newcommand{\lsim}{\lower.7ex\hbox{$\;\stackrel{\textstyle<}{\sim}\;$}}

\newcommand{\mrm}{\mathrm}

\documentclass[aps,prev,twocolumn,preprintnumbers,floatfix,nofootinbib]{revtex4-1}
\usepackage{graphicx}
\usepackage{epstopdf}
\usepackage{mathrsfs}
\usepackage{amssymb}
\usepackage{verbatim}

\def\stacksymbols #1#2#3#4{\def\theguybelow{#2}
    \def\vp{\lower#3pt}
    \def\sp{\baselineskip0pt\lineskip#4pt}
    \mathrel{\mathpalette\intermediary#1}}

\def\intermediary#1#2{\vp\vbox{\sp
     \everycr={}\tabskip0pt
     \halign{$\mathsurround0pt#1\hfil##\hfil$\crcr#2\crcr
              \theguybelow\crcr}}}

\def\be{\begin{equation}}
\def\ee{\end{equation}}
\def\bea{\begin{eqnarray}}
\def\eea{\end{eqnarray}}

\def\m{\mu}

\def\sp{\;\;\;,\;\;\;}

\def\mrm{\mathrm}

\def\lsim{\raise0.3ex\hbox{$\;<$\kern-0.75em\raise-1.1ex\hbox{$\sim\;$}}}
\def\gsim{\raise0.3ex\hbox{$\;>$\kern-0.75em\raise-1.1ex\hbox{$\sim\;$}}}

\def\inbar{\,\vrule height1.5ex width.4pt depth0pt}

\def\IC{\relax\hbox{$\inbar\kern-.3em{\rm C}$}}
\def\IQ{\relax\hbox{$\inbar\kern-.3em{\rm Q}$}}
\def\IR{\relax{\rm I\kern-.18em R}}
 \font\cmss=cmss10 \font\cmsss=cmss10 at 7pt
\def\IZ{\relax\ifmmode\mathchoice
 {\hbox{\cmss Z\kern-.4em Z}}{\hbox{\cmss Z\kern-.4em Z}}
 {\lower.9pt\hbox{\cmsss Z\kern-.4em Z}}
 {\lower1.2pt\hbox{\cmsss Z\kern-.4em Z}}\else{\cmss Z\kern-.4em Z}\fi}

\def\comment#1{}

\def\u1x{U(1)_X}
\newcommand{\nc}{\newcommand}
\nc{\LL}{L}
\nc{\vv}{\tilde{v}}
\nc{\ccdot}{\!\cdot\!}
\nc{\gsm}{G_{SM}}
\nc{\vfive}{\mathbf{5}\oplus\mathbf{\overline{5}}}
\nc{\vten}{\mathbf{10}\oplus\mathbf{\overline{10}}}
\nc{\zhol}{Z^{\rm hol}}
\nc{\xfb}{\,{\rm fb}}

\begin{document}

%
%

\preprint{LPT--Orsay 11/66}

\title{Higgs searches and singlet scalar dark matter: Combined constraints  from XENON100 and the LHC}

\author{Yann Mambrini$^{a}$}
\email{Yann.Mambrini@th.u-psud.fr}

\vspace{0.2cm}
\affiliation{
${}^a$ Laboratoire de Physique Th\'eorique 
Universit\'e Paris-Sud, F-91405 Orsay, France}

\begin{abstract}

XENON100 and the LHC are two of the most promising machines to test the physics beyond the Standard Model.
In the meantime, indirect hints push us to believe that the dark matter and Higgs boson could be the two next 
fundamental particles to be discovered.
Whereas ATLAS and CMS have just released their new limits on the Higgs searches, XENON100 obtained very recently strong constraints
on DM-proton elastic scattering. 
In this work, we show that when we combined WMAP and the most recent results of XENON100, the invisible width of the Higgs
to scalar dark matter is negligible($\lesssim 10 \%$), 
except in a small region with very light dark matter ($\lesssim 10$ GeV) not yet excluded by XENON100 or around 60 GeV
where the ratio can reach 50\% to 60\%. 
The new results released by the Higgs searches of ATLAS and CMS set 
very strong limits on the elastic scattering cross section, even restricting it to the region
$8 \times 10^{-46} \mrm{cm^2}  \lesssim \sigma_{S-p}^{SI}\lesssim 2 \times 10^{-45} \mrm{cm^{2}}$  in the hypothesis
 $135 \mrm{~GeV} \lesssim M_H \lesssim 155 \mrm{~GeV}$. 
\end{abstract}

\maketitle


\maketitle


\setcounter{equation}{0}



\section{Introduction}

Two of the most important issues in particle physics phenomenology are the nature
 of the dark matter and the mechanism to
realize spontaneously the electroweak symmetry breaking of the Standard Model (SM).
 The observations made
by the WMAP collaboration \cite{WMAP} show that the matter content of the universe is dark,
making up about 85 \% of the total amount of matter whereas
 the XENON collaboration recently released its constraints
on direct detection of Dark Matter \cite{Aprile:2011ts}. These constraints are the most 
stringent in the field nowadays, and begin to exclude a significant part of the parameter space
of the Weakly Interacting Massive Particle (WIMP) paradigm. 
In the meantime, the accelerator collaborations
ATLAS \cite{ATLAS}, CMS \cite{CMS} and D0/CDF \cite{D0, TEVATRON} presented 
their results concerning the Higgs searches. It is obvious that the Higgs hunting
at LHC is intimately linked with measurement of elastic scattering on nucleon,
especially in Higgs-portal like models where the Higgs boson is the key particle
exchanged through annihilation/scattering processes.
It has already been showed recently that a combined LEP/TEVATRON/XENON/WMAP analysis
can restrict severely the parameter space allowed in generic constructions \cite{Mambrini:2011pw}.
In this work, we apply such analysis in the specific context of a scalar singlet dark matter extension
of the Standard Model and show that most of the region allowed by WMAP
will be excluded/probed by LHC and XENON100 by the end of next year.

\noindent
The paper is organized as follows:
we summarize in section II the scalar singlet extension of the Standard Model and study its direct 
detection modes based on
recent analysis of the nucleon structure and their influences on the detection prospects. We then devote
section III to the invisible branching ratio of the Higgs.
We show that after combining WMAP and the last XENON100 constraints, the invisible width of the Higgs
is negligible,  making it a SM Higgs for which ATLAS and CMS observability studies can be applied.
We then include in section IV the new LHC/TEVATRON analyses released very recently and show that a large
part of the parameter space of the model is already excluded. We then concentrate in section V on the direct detection cross section
one can expect if a Higgs boson mass $M_H \simeq 145$ GeV is observed in a near future.
We then conclude in section  VI.

\section{Direct detection and nucleon structure}

\subsection{The model}

The simplest extension of the SM is the addition of a real singlet scalar field.Although it is possible to generalize to scenarios with more than one singlet, the simplest
case of a single additional singlet scalar provides a useful framework to analyze
the generic implications of an augmented scalar sector to the SM.
The most general renormalizable potential involving the SM Higgs doublet $H$ and the
singlet $S$ is

\bea
{\cal L}&=& {\cal L}_{SM} + (D_{\mu}H)^\dag (D^{\mu}H) + \frac{1}{2} \mu_H^2 H^\dag H 
-\frac{1}{4}\lambda_H H^4
\nonumber
\\
&+& \frac{1}{2} \partial_{\mu}S \partial^{\mu}S
-\frac{\lambda_{S}}{4} S^4 - \frac{\mu_S^2}{2} S^2 - \frac{\lambda_{HS}}{4}S^2 H^{\dag}H
\nonumber
\\
&-&\frac{\kappa_1}{2} H^\dag H S - \frac{\kappa_3}{3}S^3 - V_0
\label{Eq:Lagrangian}
\eea 
 
 \noindent
 where $D_{\mu}$ represents the covariant derivative.
 We have eliminated a possible linear term in $S$ by a constant shift,
absorbing the resulting S-independent term in the vacuum energy $V_0$. 
We require that the minimum of the potential occur at $v =$ 246 GeV . Fluctuations
around this vacuum expectation value are the SM Higgs boson.
 For the case of interest here for which $S$ is stable and may be a dark matter candidate,
we impose a $Z_2$ symmetry on the model, thereby eliminating the $\kappa_1$ and $\kappa_3$  terms. 
 We also require that the true vacuum of the theory satisfies $\langle S \rangle=0$, thereby precluding
  mixing of $S$ and the SM Higgs boson and the existence of cosmologically problematic domain walls. In
this case, the masses of the scalars are

 \be
 M_H = \sqrt{2} \mu_H ~~~~~~~~ m_S= \sqrt{\mu_S^2 + \frac{\lambda_{HS}}{\lambda_H}\frac{\mu_H^2}{2}}
 \ee
 
 \noindent
 and the HSS coupling generated is 
 
 \be
{\cal C}_{HSS} : ~~~-\frac{\lambda_{HS}M_W}{2 g}.
 \ee

\begin{figure}
    \begin{center}
   \includegraphics[width=3.3in]{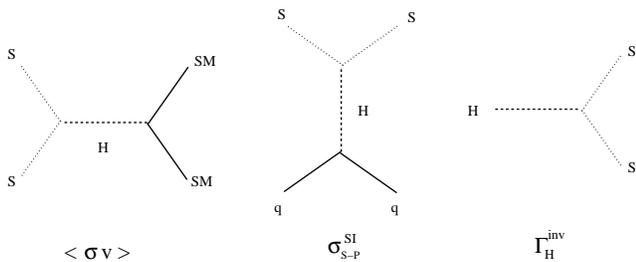}
   
          \caption{{\footnotesize
Feynman diagram for the annihilation cross section (left) direct detection scattering (center) and invisible
width of the Higgs (right).
}}
\label{Fig:Feynman}
\end{center}
\end{figure}

 \noindent
 We show in Fig.\ref{Fig:Feynman} the Feynman diagrams relevant for our analysis\footnote{The
quartic coupling SSHH which can be efficient in the computation of the relic abundance if
 $m_s \gtrsim M_H$ is also present. We obviously took it into account
in our numerical analysis but its contribution to the annihilation processes is always subdominant.}
.
Different aspects of scalar singlet extension of the SM has already been studied in 
\cite{McDonald:1993ex,McDonald:2001vt,Burgess:2000yq,Patt:2006fw,Meissner:2006zh,
Davoudiasl:2004be,Zhu:2006qx,He:2008qm,Kanemura:2010sh,
Aoki:2009pf,Guo:2010hq,Barger:2010mc, Espinosa:2007qk,SungCheon:2007nw}
whereas a nice preliminary analysis of its dark matter consequences can be found in 
\cite{Barger:2007im}. Some authors also tried to explain the DAMA and/or COGENT 
excess \cite{Andreas:2008xy,Andreas:2010dz,Tytgat:2010bt}
whereas other authors probed the model by indirect searches \cite{Yaguna:2008hd,Goudelis:2009zz,Arina:2010rb},
or looked at the consequences of earlier XENON data \cite{Cai:2011kb,Biswas:2011td,Farina:2011bh}

 \subsection{The nucleon structure uncertainties}

Since several years it is known
 that the uncertainties generated
by the quark contents of the nucleons can be as important (if not more) than astrophysical uncertainties.
Some authors pointed out this issue and applied it to supersymmetric models \cite{Ellis:2008hf,Giedt:2009mr},
in effective operator approach \cite{Gao:2011ka}
or even in the scalar extension of the SM \cite{Andreas:2008xy}, but rarely taking into 
account the latest lattice results \cite{Giedt:2009mr}. Indeed, due to its large Yukawa coupling, the strange quark
and its content in the nucleon is of particular interest in the elastic scattering of the dark matter on the proton.
The spin independent part of the cross section can be written 

\be
\sigma^{SI}_{S-N} = \frac{4 m_r^2}{\pi} \left[   Z f_p +(A-Z) f_n \right]^2
\ee

\noindent
where $m_r = m_N m_S/(m_S + m_N)$ is the $S-$nuclear reduced mass and 

\be
f_N =  m_N  \left( \sum_{q=u,d,s} f_{q}^N \frac{{\cal A}_q}{m_q} + \frac{2}{27} f^N_{H} \sum_{q=c,b,t} 
\frac{{\cal A}_q}{m_q} \right) 
\ee

\noindent
with ${\cal A}_q$ the scattering amplitude on a single quark $q$ and
$f^N_q= (m_q/m_N) \langle N | \bar q q |N\rangle$ is the reduced (dimensionless) sigma terms
of the nucleon N, and $f^N_H = 1 - \sum_{q=u,d,s} f^N_q $ \cite{Shifman:1978zn,Vainshtein:1980ea}.

There are different ways of extracting the reduced dimensionless nucleon (N) sigma terms 
$f^N_q \equiv (m_q/m_N) \langle N | \bar q q | p \rangle$. 
This sigma terms can be derived by phenomenological estimates of the $\pi-N$ scattering $\Sigma_{\pi N}$
(see \cite{Young:2009ps} and references therein for a review): 
$\Sigma_{\pi N} \equiv m_N f_l=m_l \langle N | \bar u u + \bar d d | N \rangle$ with
$m_l=(m_u+m_d)/2$. While an early experimental extraction \cite{Koch:1982pu} gave 
$\Sigma_{\pi N} = 45 \pm 8$ MeV, a more recent determination \cite{Pavan:2001wz} obtained 
$\Sigma_{\pi N} = 64 \pm 7$ MeV.

\begin{figure}
    \begin{center}
   \includegraphics[width=3.5in]{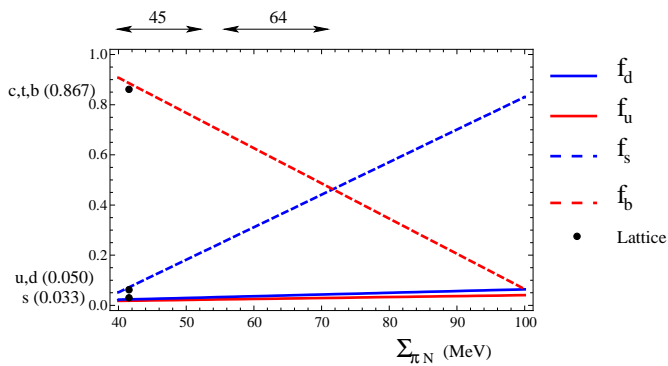}
   
          \caption{{\footnotesize
 Sigma commutators of the proton with two different phenomenological measurements of $\Sigma_{\pi N}=45 \pm 8$ MeV 
 \cite{Koch:1982pu} and $64 \pm 7$ MeV \cite{Pavan:2001wz}. We also showed the mean evaluation
 from more recent lattice results \cite{Young:2009zb} (left).
}}
\label{Fig:fTu}
\end{center}
\end{figure}

 On the other hand, the study of the breaking of
$SU(3)$ within the baryon octet and the observation of the spectrum leads to derive a constraint on the
non-singlet combination $\sigma_0 = m_l  \langle N | \bar u u + \bar d d - 2 \bar s s |N \rangle$. 
Chiral effective field theory lead to a value
$\sigma_0 = 36 \pm 7 ~ \mrm{MeV}$.
Following \cite{Cheng} by introducing $z=(\langle N | \bar u u + \bar s s | N \rangle)/(\langle N | \bar d d + \bar s s | N \rangle)=1.49$
one obtains

\bea
&&
f_d = \frac{m_d}{m_N} \frac{\Sigma_{\pi N}}{m_u+m_d} \frac{y(z-1)+2}{1+z}
\nonumber
\\
&&
f_u = \frac{m_u}{m_N} \frac{\Sigma_{\pi N}}{m_u+m_d} \frac{y(1-z)+2z}{1+z}
\nonumber
\\
&&
f_s = \frac{m_s}{m_N} \frac{\Sigma_{\pi N}}{m_u+m_d} ~~y
\eea

 \noindent
 where $y=2 \langle N| \bar s s |N\rangle / \langle N | \bar u u + \bar d d |N \rangle = 1-\sigma_0/\Sigma_{\pi N}$ represents the strange fraction
 in the nucleon. We show in Fig.\ref{Fig:fTu} the dependance of $f_q$ as function of $\Sigma_{\pi N}$.
 The two extreme values are obtained with the lower bound of $\Sigma_{\pi N}$ at $1\sigma$ extracted from \cite{Koch:1982pu}
 (37 MeV)  and the higher bounds from \cite{Pavan:2001wz} (71 MeV) which gives
 for ($m_u,m_d,m_s,m_c,m_b,m_t,m_p)=(2.76,5.,94.5,1250,4200,171400,938.3)~\mrm{[MeV]}$ :    
 \bea
 &&
 f_u^{min } = 0.016  ~~~~~ f_u^{max}= 0.030
 \nonumber
 \\
 &&
  f_d^{min } = 0.020  ~~~~~ f_d^{max}= 0.044
  \nonumber
  \\
  &&
  f_s^{min } = 0.013  ~~~~~ f_s^{max}=0.454
  \label{Eq:sigpheno}
 \eea

These limitations on the phenomenological estimation of  the strange structure of the nucleon clearly open
the way for lattice QCD to offer significant improvement. Using the Feynman--Hellman relation 
$ f_q = (m_q/M_N)\partial M_N/ \partial m_q$ different authors have extracted the light--quark
and strangeness sigma terms (see \cite{Young:2009ps} for a clear review). The last results obtained by the MILC collaboration
\cite{Toussaint:2009pz} and by the authors of (labeled "Young" from now on)
\cite{Young:2009zb}  provide stringent new limits on the strange quark 
sigma-terms. The modern lattice results for $f_s$ agree that the size is substantially smaller than
has been previously thought :

\be
f_s^{\mrm{Young}} =  0.033 \pm 0.022~~~~~    f_s^{\mrm{MILC}}=  0.069 \pm 0.016
\ee

\noindent
These two results are marginally consistent, although there may be differences in how the derivative
with respect to $m_s$ is taken. Moreover, they tend to favor the smaller phenomenological evaluation
of $\Sigma_{\pi N}$.  
In the following, we will consider the central values of $f_q$ extracted from the Young et al analysis and referred it
to the "lattice" one : $f_u=f_d=f_l=0.050$, and $f_s=0.033$, and the maximum and minimum
values for $f_q$ given by phenomenological references \cite{Koch:1982pu} and \cite{Pavan:2001wz}
(Eq.\ref{Eq:sigpheno}).
We adapted the code micrOMEGAs \cite{Micromegas} to the different values of $f_q$ depending
on the model we used, and modified it to include the new couplings/spectrum/interactions induced by the singlet scalar
extension of the SM\footnote{We want to thank warmly S. Pukhov for his help to solve technical problems
related to the code.}.

\begin{figure}
    \begin{center}
   \includegraphics[width=3.5in]{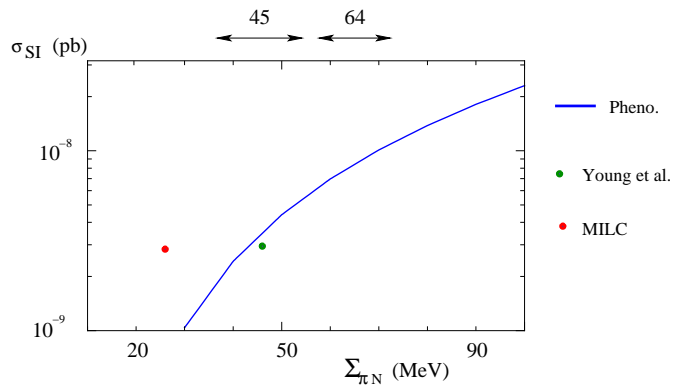}
   
          \caption{{\footnotesize
Spin independent elastic scattering cross section as function of the pion-nucleon sigma-term $\Sigma_{\pi N}$
for a scalar dark matter respecting WMAP constraint:
 $m_s=90$ GeV, $\lambda_{HS}=0.2$, $m_h=130$ GeV giving $\Omega_S h^2=0.102$. We also represented
 the central values of the cross section for the lattice simulations \cite{Toussaint:2009pz} and 
 \cite{Young:2009zb} labelled "MILC" and "Young"
 respectively.
}}
\label{Fig:SigfTs}
\end{center}
\end{figure}

\noindent 
As we can see in Fig.\ref{Fig:SigfTs}, these uncertainties have a strong impact on the direct detection cross section,
up to one order of magnitude. 
We also plotted the value of $\sigma_{SI}$  obtained by the two lattice groups that we took into consideration
in our analysis, corresponding to the central values ($\Sigma_{\pi N}, \sigma_{SI}$) =
($26$ MeV, $2.84 \times 10^{-9}$ pb) \cite{Toussaint:2009pz} and ($47$ MeV; $2.95 \times 10^{-9}$ pb) \cite{Young:2009zb}.
We clearly see that the lattice results are in much more accordance with the lower bound on $\Sigma_{\pi N}$:
$\sigma_{SI}^{min}(\Sigma_{\pi N}= 37 ~\mrm{MeV})=1.93 \times 10^{-9}$ pb, whereas
 $\sigma_{SI}^{max}(\Sigma_{\pi N}=71 ~\mrm{MeV})=1.05 \times 10^{-8}$ pb. We compiled all the necessary values
 of $f_i$ in the following table
 
 \begin{center}
 \begin{tabular}{|c|c|c|c|}
\hline
$f_i$ &Lattice & Min & Max \\
 \hline
 $f_u$ & 0.050 & 0.016& 0.030  \\
$f_d$ & 0.050 &0.020 & 0.044 \\
 $f_s$ & 0.033 & 0.012& 0.454 \\
$f_{c,t,b}$ &0.867 &0.952 & 0.472 \\
$f=\sum f_l + 3 \times \frac{2}{27} f_H$ &0.326 &0.260 & 0.629 \\
\hline
\end{tabular}
\end{center}

 \noindent
In the rest of the paper, we will always present our results with the evaluation
of $f_s$ given by the maximum and minimum allowed value for $\Sigma_{\pi N}$ and the lattice extraction
of Young et al. (which gives quite similar cross section to the one obtained by the MILC group as we concluded above).

\begin{figure}
    \begin{center}
    \hspace{-1.cm}
   \includegraphics[width=1.8in]{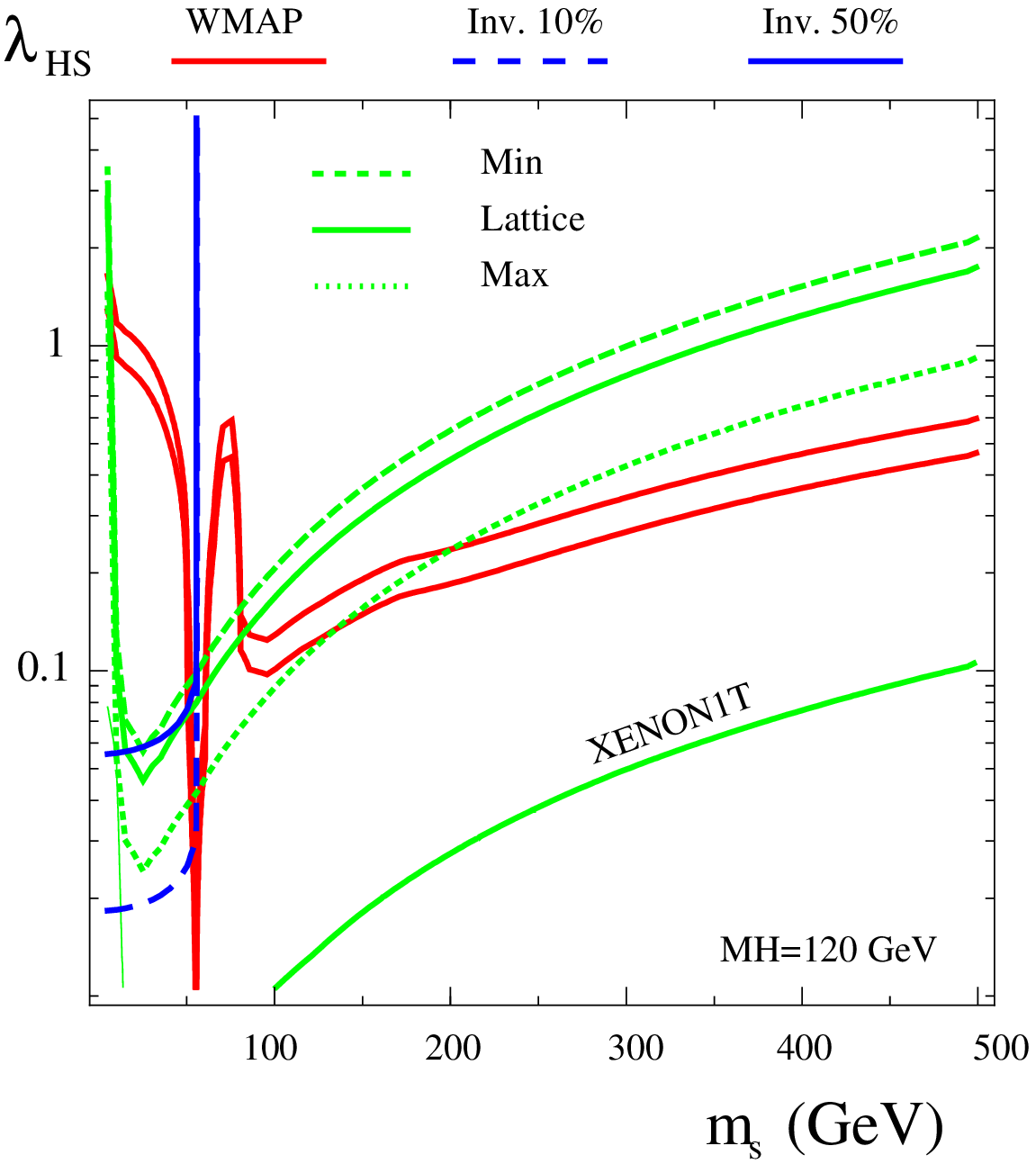}
   \includegraphics[width=1.8in]{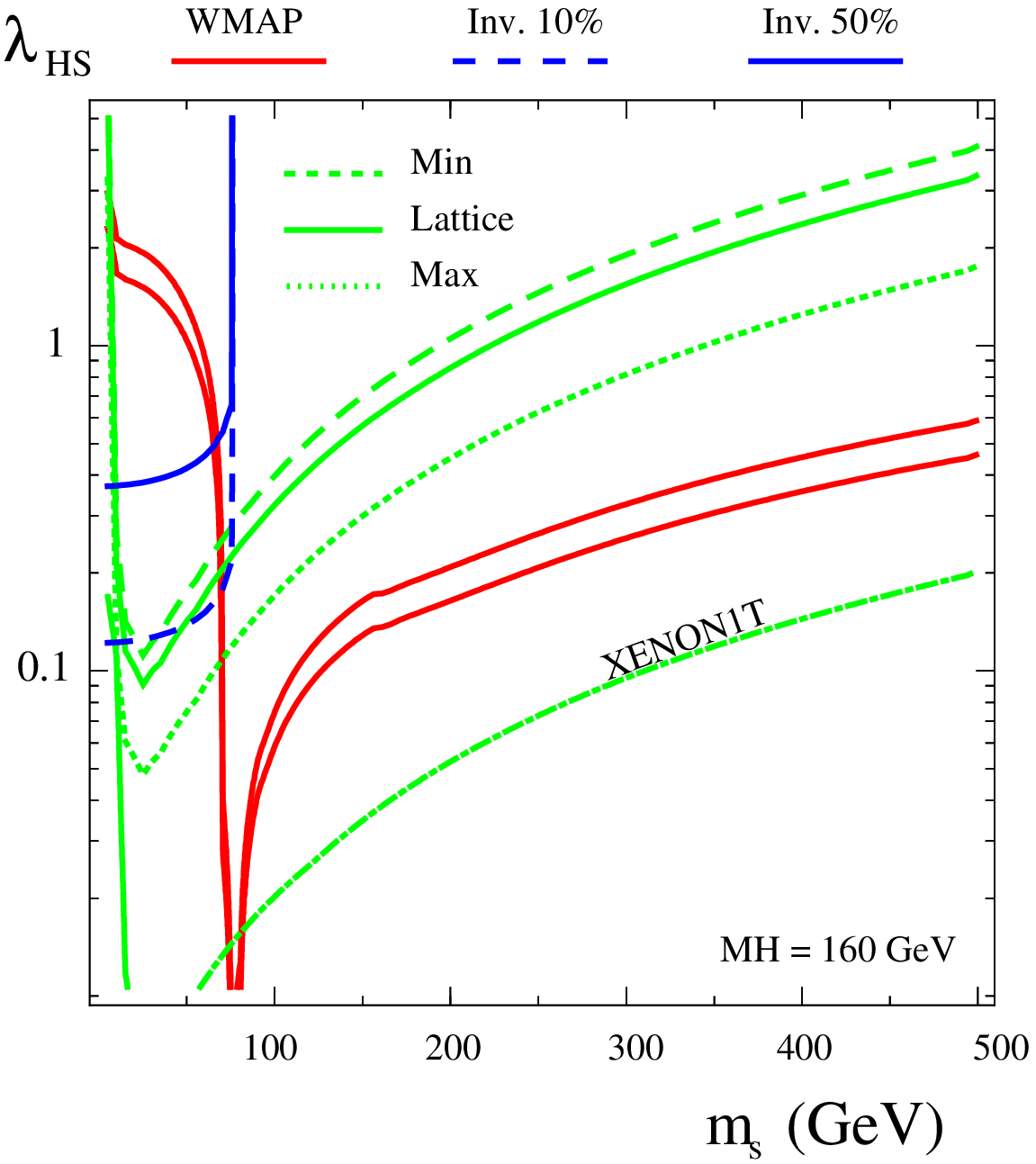}
   
      \hspace{-1.cm}
    \includegraphics[width=1.8in]{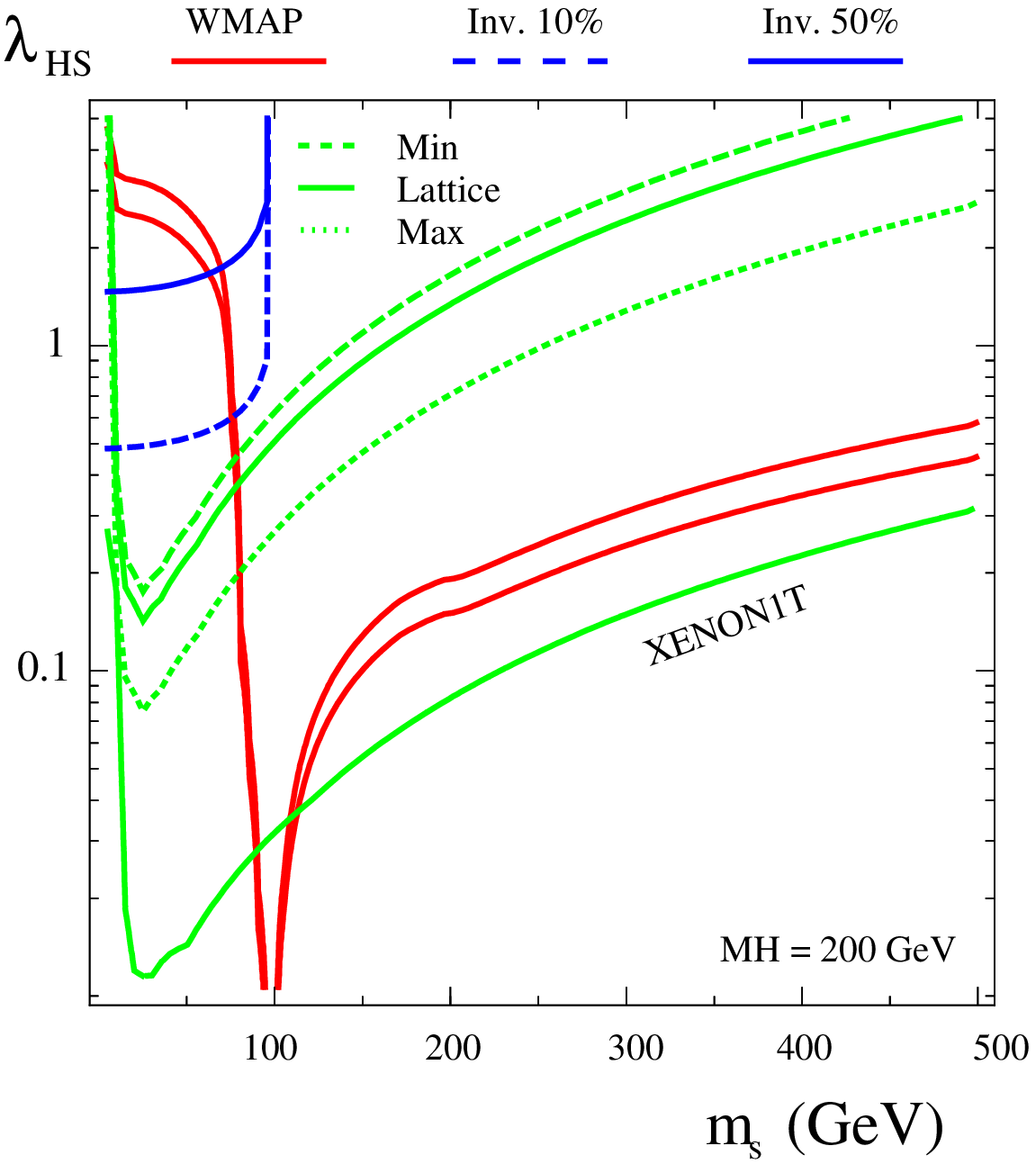}
            \includegraphics[width=1.8in]{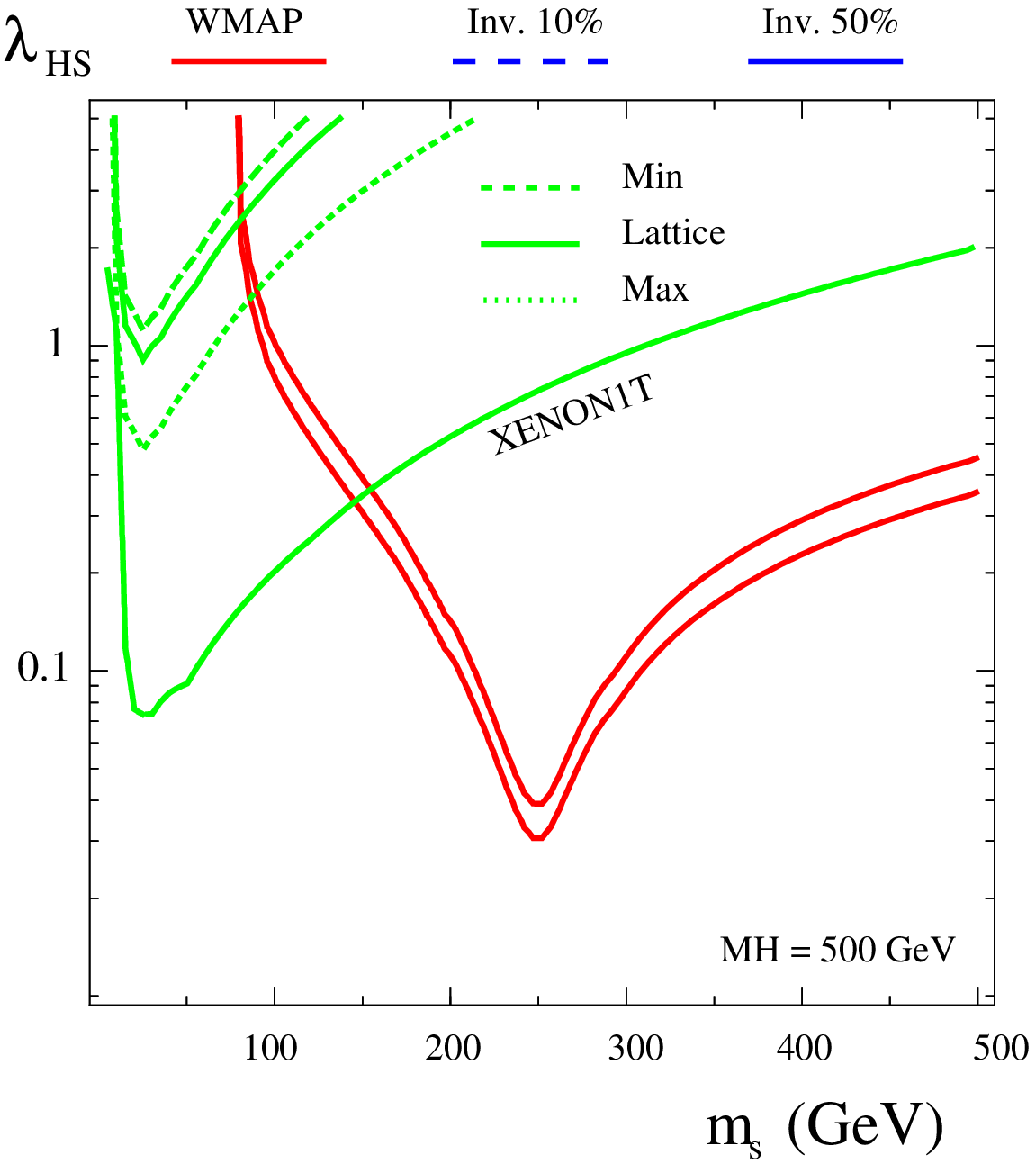}
    
          \caption{{\footnotesize
Parameter space allowed in the plane ($\m_S,\lambda_{HS}$) for different Higgs masses (120, 160, 200 and 500 GeV)
taking into account the last XENON100 data and the XENON 1T projection, with different values for the strange
structure of the nucleon. We also show the invisible branching fraction of the Higgs boson width (10 and 50 \% respectively).
See the text for details.
}}
\label{Fig:Scan}
\end{center}
\end{figure}

\noindent
We show in Fig.\ref{Fig:Scan} the influence of the sigma-terms in the region excluded/accessible by XENON100
for different Higgs masses.
Clearly, for the lowest values of $f_s$, constraints on the parameter space become weaker because less points 
generate a cross section exceeding the direct detection bounds. We also represented the expectation
of a XENON 1T experiment and showed that it would reach 80 \% of the WMAP allowed parameter space, 
but could not tell anything for a heavy Higgs boson $M_H \gtrsim 300$ GeV, which is complementary to the LHC searches:
 due to the specific decay modes of the Higgs boson, LHC is more sensitive to a heavy
 Higgs than to a light one.

\section{The invisible Higgs width : a XENON100/LHC complementarity}

Recently the XENON100 collaboration released new data, the most stringent in the field
of Dark Matter detection\footnote{Keeping an eye on the results of COGENT collaborations \cite{Aalseth:2010vx}, 
recent works showed there exists a tension between XENON100 and COGENT \cite{Schwetz:2011xm}, 
or not \cite{Hooper:2011hd}. We thus safely 
decided not to discuss in detail the COGENT issue in our analysis} \cite{Aprile:2011ts}.
Moreover, recently CRESST experiment released their analysis in the low mass region \cite{Angloher:2011uu} 
and seems to converge with DAMA/LIBRA and CoGENT toward a possible light dark matter signal for a mass
aound 10 GeV \cite{Kelso:2011gd}. In the meantime,
if $m_S \lesssim M_H/2$ the invisible width decay\footnote{See the works in \cite{Hinv} for an earlier study of
invisible width of the Higgs. Moreover, during the revision of this study, several independant work confirming 
our results were published
in \cite{Raidal:2011xk} and \cite{He:2011de}} of the Higgs $H \rightarrow SS$ could 
perturbate the Higgs searches at LHC based on SM Higgs branching ratio (see Eq.\ref{Eq:HiggsWidth}
 and \cite{Abdel} for a review on the SM Higgs width computation). 
However, one can easily understand that there exists a tension between the direct detection measurement
and the invisible branching ratio. Indeed, for decreasing mass of DM ($m_S \lesssim 100$ GeV), 
the spin independent cross section increases. We show in Fig.\ref{Fig:Scan} the regions in the 
($m_S$; $\lambda_{HS}$) plane where the invisible branching fraction reaches 10 and 50 percents 
(dashed and full blue lines),  for different values of the Higgs mass.
 As we noticed in the previous section, one  needs a low value for $\lambda_{HS}$ to respect 
 the stringent XENON100 bounds.
  This is precisely in this regime  ($m_s \lesssim M_H/2$) that the invisible width could interfere in the Higgs searches.
 However, the low value of $\lambda_{HS}$ restricted by XENON100 made this branching ratio very small. 
Quantitatively speaking, one needs to compare the invisible Higgs width ($H \rightarrow SS$)

\be
\Gamma^{{\mrm inv}}_H= \frac{\lambda_{HS}^2 M_W^2}{32 \pi g^2 M_H^2} \sqrt{M_H^2 - 4 m_S^2}
\label{Eq:HiggsWidth}
\ee

\noindent
with the spin independent scattering cross section on the proton

\be
\sigma^{SI}_{S-p} = \frac{m_p^4 \lambda_{HS}^2 (\sum_{q} f_q )^2}{16 \pi (m_p + m_S)^2 M_H^4}
\label{Eq:SigmaSI}
\ee

\noindent
Combining Eq.(\ref{Eq:HiggsWidth}) and Eq.(\ref{Eq:SigmaSI}) one obtains

\be
\frac{\Gamma_H^{\mrm Inv}}{\sigma^{SI}_{S-p}}= \frac{(m_S+m_p)^2 M_H^2 M_W^2 \sqrt{M_H^2-4 m_S^2}}{2 g^2 f^2 m_p^4}
\ee

\noindent
which reaches its maximum for $m_s-S^{max}= ( -m_p + \sqrt{m_p^2 + 6 M_H^2}) /6\simeq 65.2 $ GeV for $M_H=160$ GeV for instance.
We can then compute the maximum value of the invisible width of the Higgs as a function of the scattering cross section on the
proton :

\be
\Gamma_{H,max}^{inv} =  \frac{(m_S^{max}+m_p)^2 M_H^2 M_W^2 \sqrt{M_H^2-4 (m_S^{max})^2}}{2 g^2 f^2 m_p^4}  \sigma^{SI}_{S-p}
\label{Eq:GammainvSigmaSI}
\ee

\begin{figure}
    \begin{center}
    \hspace{-0.5cm}
   \includegraphics[width=3.5in]{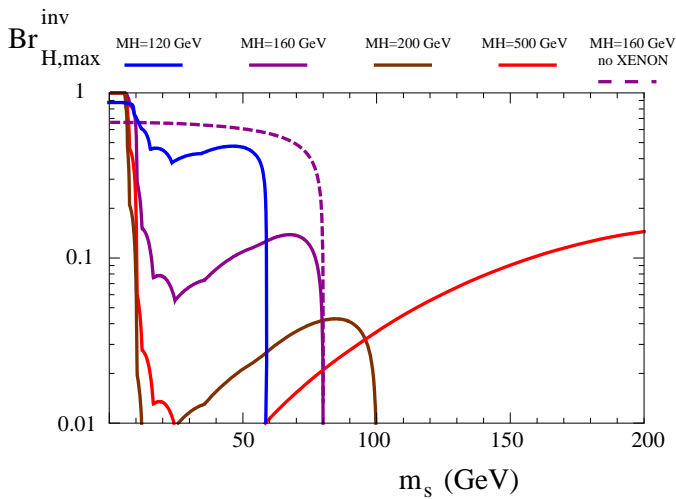}
   
          \caption{{\footnotesize
Maximum Higgs invisible branching ratio as a function of the dark matter mass for different higgs masses taking into
account the last XENON100 constraint. We also show an example of invisible branching ratio for $M_H=160$ GeV
before taking into account XENON100 constraint.
}}
\label{Fig:BrInv}
\end{center}
\end{figure}

\noindent
We show in Fig.\ref{Fig:BrInv} the value of the maximal branching ratio as function of the dark matter mass for
different values of $M_H$, taking into account the maximum value ($m_S-$dependent) of $\sigma_{S-p}^{SI}$
allowed by the last data released by XENON100. We see that the XENON100 constraints impose
a very low invisible Higgs branching ratio. To illustrate it, we also plotted a typical example of branching
ratio for $M_H=160$ GeV $without$ taking into account the XENON100 data (dashed magenta). 
Whereas 80\% of the Higgs could decay invisibly, after applying the XENON constraint on $\sigma_{S-p}^{SI}$
its invisible branching fraction reaches only 10 \% {\it at its maximum} (corresponding to $m_S=m_S^{max}$).
Only for a very light Higgs ($M_H \lesssim 120$ GeV) $Br^{inv}_H$ can reach 50 \%.
The analysis was run with the value of the sigma terms of the nucleon given by the Young et al. analysis
\cite{Young:2009zb}. In fact, this choice is very conservative because if we took values of $f_i$ corresponding
to the maximum (unphysical) value of $\Sigma_{\pi N}$, one would obtain even a lower invisible width
for the Higgs boson (Eq.\ref{Eq:GammainvSigmaSI}).

On top of that, when we include the WMAP constraint the allowed region shrinks and mainly branching fractions less than 
$\simeq10^{-1}$ still resist to all the constraints.
However, some points
with high invisible width still survive. They correspond
to two distinct regions: 

\begin{itemize}
\item{ A region with very light scalar ($m_S \lesssim 10$ GeV) still not yet excluded by the precision
of XENON100 experiments due to its high threshold. This correspond to very large invisible branching 
ratio}

\item{A region with $50 \mrm{GeV} \lesssim m_S \lesssim 70$ GeV with branching ratio which can
 reach 60\% to 70 \% which is the region taken in consideration in \cite{Raidal:2011xk}}.

\end{itemize} 

\begin{figure}
    \begin{center}
   \includegraphics[width=3.5in]{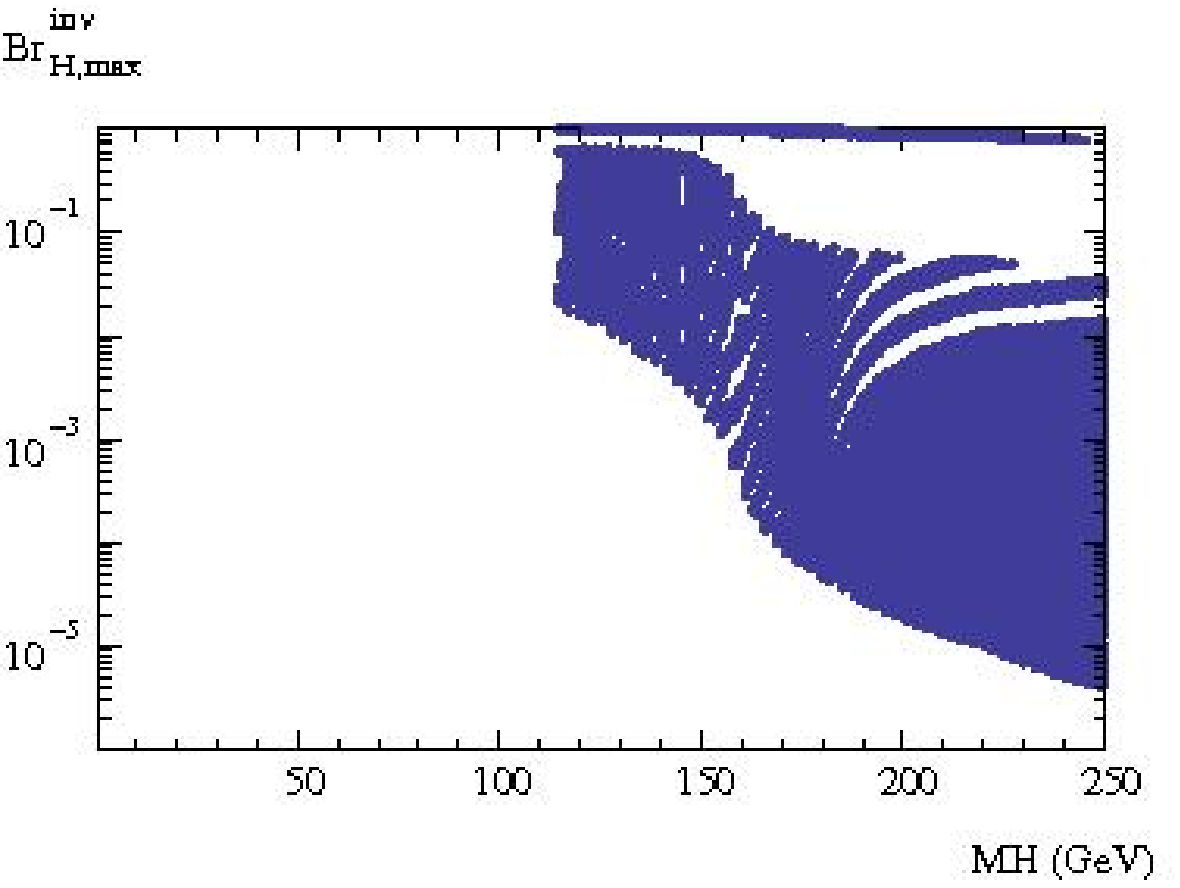}
   
      \includegraphics[width=3.5in]{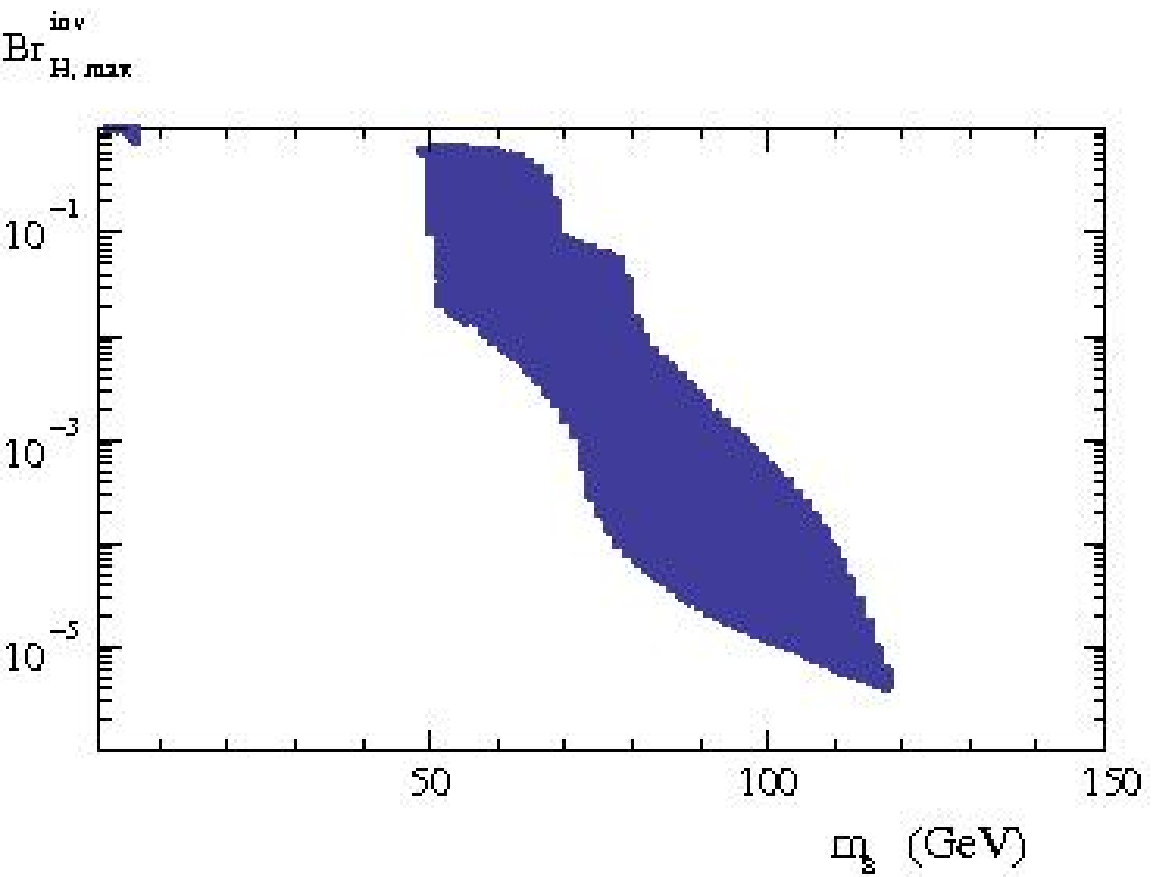}
   
          \caption{{\footnotesize
Maximum Higgs invisible branching ratio as a function of the Higgs mass (top) and the dark matter mass (bottom)
 after a complete 
scan on $\lambda_{HS}, m_s$ and  $M_H$ taking into account the constraint on WMAP and  applying the
 last XENON100 results. We clearly see that the points with high invisible Higgs branching ratio are limited to a region
 with very low dark matter mass ($\lesssim 10$ GeV) and can reach $\sim 50$ \% for masses around 60 GeV
 as was noticed by \cite{Raidal:2011xk}
}}
\label{Fig:BrInvScan}
\end{center}
\end{figure}

\noindent
We show the effects of combining WMAP and XENON100 data in Fig.\ref{Fig:BrInvScan}. 
As one can see, except these two particular regions,  the majority of points respecting WMAP and XENON100 constraints give 
 very low invisible width. As a conclusion, we can 
affirm that the Higgs searches at LHC with a scalar dark matter is not affected:
the behavior of the Higgs is a Standard Model one, even including a singlet in the game. This is one of the
strongest conclusions of this work, and the first level of complementarity between detection modes. 
It also means that we can use the standard Higgs limit searches of ATLAS and CMS and apply them in the model. 
They are only slightly affected by the presence of the scalar dark matter. However, in our numerical study, we
obviously took into consideration the invisible Higgs width to apply the CMS and ATLAS constraints.

Due to the last data released recently by CRESST collaboration \cite{Angloher:2011uu} it is interesting to notice that
some points in the parameter space around $m_S \simeq$ 10 GeV are not yet excluded by the latest XENON100
constraints as can be seen in the upper left corner of Fig.\ref{Fig:BrInvScan} (bottom). These points generates
a Higgs completely invisible at the LHC.
This corresponds to the region near Br($H \rightarrow SS$) $\simeq$ 100\% in Fig.\ref{Fig:BrInvScan}  (top).

\section{The Higgs Hunting : an LHC/XENON100 complementarity}

\begin{figure}
    \begin{center}
      \hspace{-0.5cm}
   \includegraphics[width=1.7in]{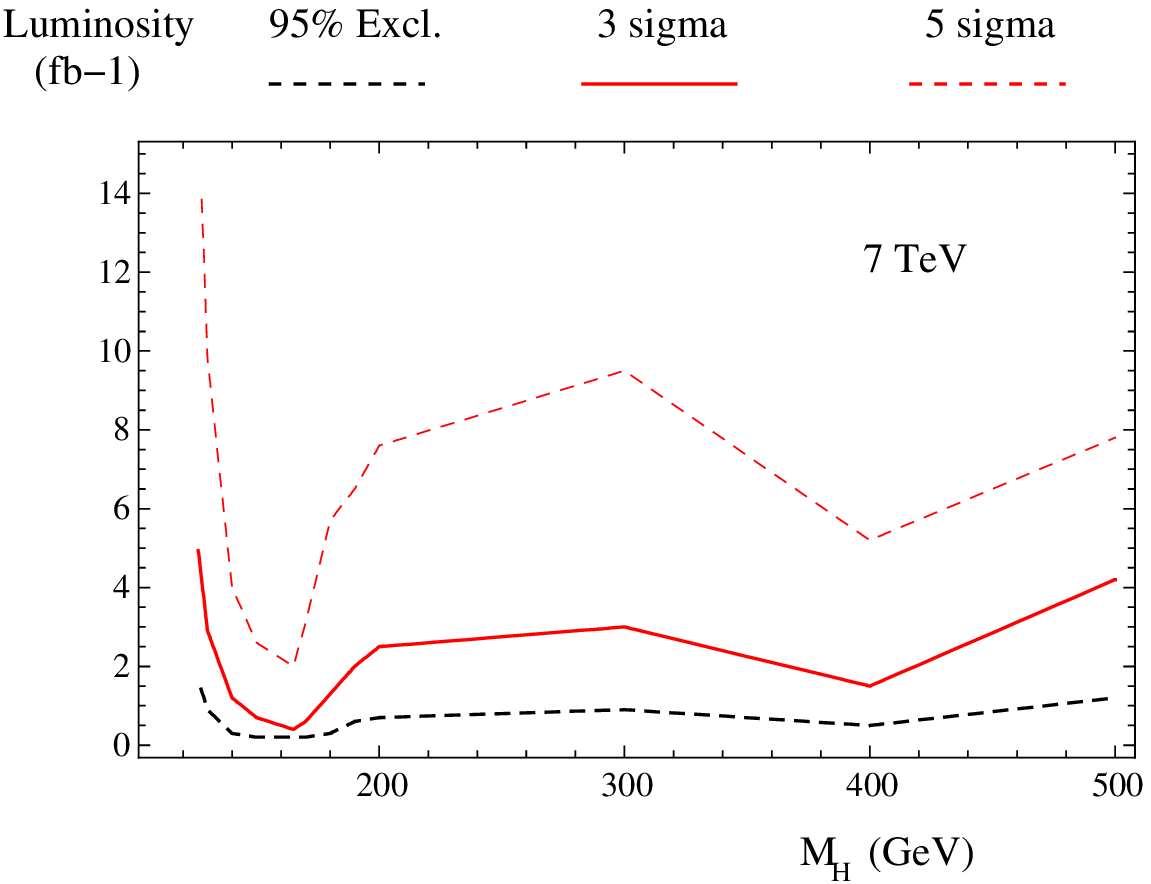}
      \includegraphics[width=1.7in]{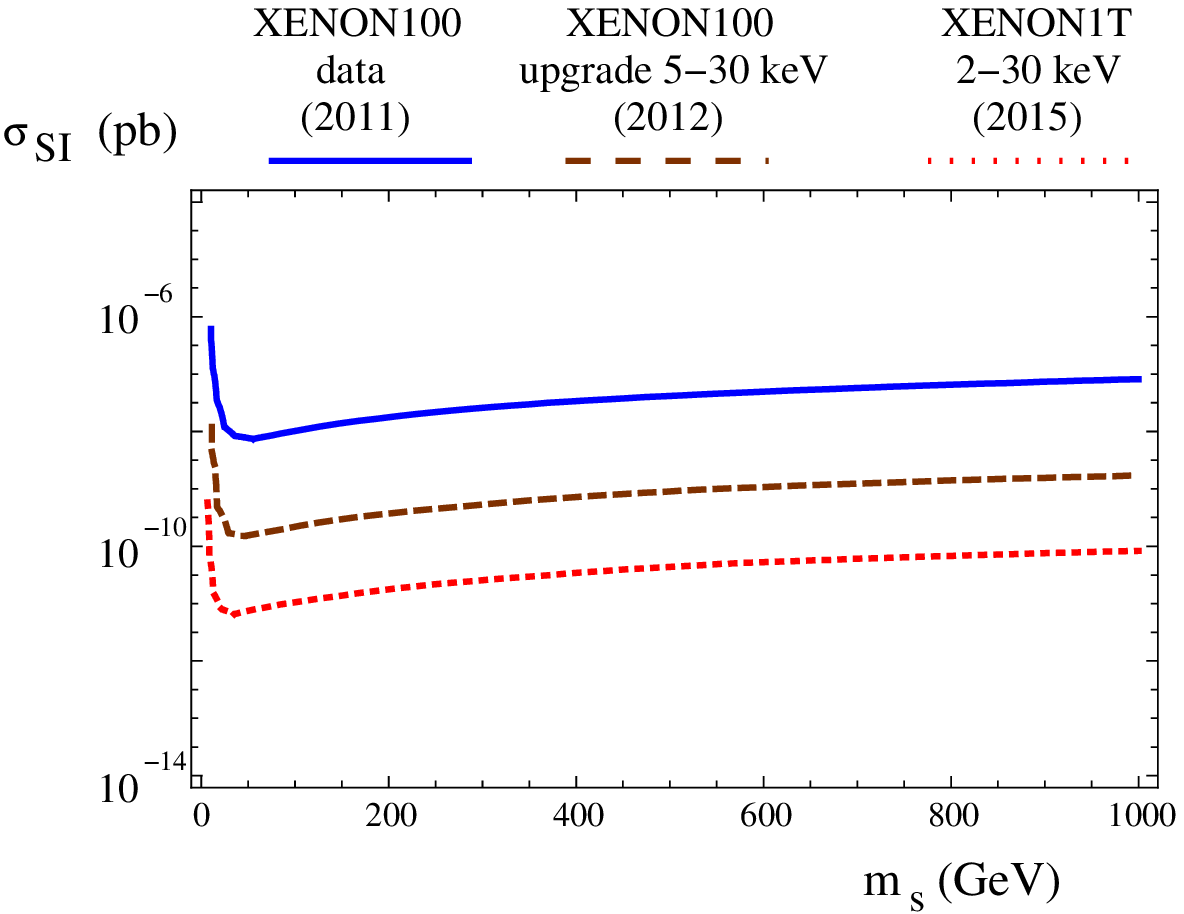}
          \caption{{\footnotesize
          Luminosity required to give exclusion (95 \% CL, dashed black), evidence ($3 \sigma$ full red) or discovery (5$\sigma$  dashed red)
          sensitivity for a SM Higgs \cite{LumiATLAS} with data at $\sqrt{s}$ = 7  TeV (left);
          Spin-independent  WIMP-proton cross sections limit
as a function of WIMP mass as measured by XENON100 \cite{Aprile:2011ts} (full blue), probed by different stages of the XENON program
          for 2012 (dashed brown) and 2015 (dotted red).
\cite{Aprile:2009yh}.         
}}
\label{Fig:Luminosity}
\end{center}
\end{figure}

As we observed in Fig.\ref{Fig:Scan} whereas the direct detection prospects are quite weak for
a Higgs mass $M_H \gtrsim 300$ GeV, XENON100 experiment will easily cover the region
$M_H \lesssim 130$ GeV in a near future, which is precisely the region the most difficult to reach at LHC.
In the meantime, ATLAS \cite{ATLAS} and CMS \cite{CMS} with an integrated luminosity of $1\mrm{fb^{-1}}$ had
 given at EPS \cite{EPS} their first exclusion zone.
 CMS excludes the Standard Model Higgs in the 149-206 GeV and 300-440 GeV windows, while ATLAS excludes 
 the 155-190 GeV and 295-450 GeV windows whereas the combined result given at Lepton Photon conference
 \cite{LeptonPhoton} gives the two Higgs exclusion zones $145 < M_H < 288$ and $295< M_H <  466$
 at 95\% of CL. The low mass exclusion is dominated by the search of 
 the $H \rightarrow WW \rightarrow 2l2\nu$ final state, while the high mass one is dominated by 
 $H \rightarrow ZZ$ after combining different $Z$ decay channels\footnote{TEVATRON collaborations presented
  a combined analysis \cite{TEVATRON} and mainly agree on the results obtained by ATLAS and CMS.}. 
  A summary and brief discussion of the analysis can be found in \cite{Blog}.
  Combining all this analysis and being very conservative
  one can exclude the Higgs mass between 145-288 GeV and 295-466 GeV. CMS and ATLAS
 could soon release a combined analysis closing the 288-295 windows.  We show in Fig.\ref{Fig:Luminosity} the
 luminosity required for a 
 95\% exclusion $3\sigma$ and $5\sigma$ discovery potential for ATLAS \cite{LumiATLAS}. We will use the results
 just released by ATLAS and CMS,  and project the ATLAS $5\sigma$ projection for a luminosity of $10 \mrm{fb^{-1}}$ 
 which will be the sensitivity reached by next year. We took the $5\sigma$ limit as we want to stay the more conservative 
 possible: if we supposed that the Higgs (SM Higgs as we just pointed out in the previous chapter) can be excluded, all the region 114-700 GeV
 could be excluded by the end of 2012, and thus the singlet extension of the SM.
 
 In addition to the experimental constraints on the Higgs boson mass discussed previously, there are interesting
 constraints which can be derived from assumptions on the energy range in which the SM is valid before perturbation
 theory breaks down and new phenomena should appear. These include constraints from unitarity
 in scattering amplitudes, perturbativity of the Higgs self coupling, stability of the electroweak vacuum and fine tuning.
 Whereas all the constraints bound roughly $M_H \lesssim 1$ TeV, the triviality bound which asks for perturbativity for the 
 Higgs self coupling $\lambda_H$ (Eq.\ref{Eq:Lagrangian}), one obtains from simulation of gauge theory on lattice
 a rigorous bound $M_H \lesssim 640$ GeV. This limit is in remarkable  agreement with the bound obtained by
 naively using the perturbation theory.
 Depending on the details of the cutoff scale, one can obtain an upper bound of 650 GeV \cite{Gockeler:1992zj}   or 750 GeV
 \cite{Neuberger:1992ng}. We will use a rather secured value of 700 GeV (mean value of the two results)
 through the rest of the paper\footnote{It was shown in \cite{Gonderinger:2009jp} that one can lower a little bit the
 upper bound on the Higgs mass when the scalar singlet is included in the computation of the perturbativity
 limit, but this will affect the SM bound only for a cutoff scale $\Lambda_{\mrm{cutoff}} \gtrsim 10^7$ GeV.
  To be conservative, we will 
 suppose through the study that $\Lambda_{\mrm{cutoff}} \gtrsim 1$ TeV.}.

 \begin{figure}
    \begin{center}
   \includegraphics[width=3.5in]{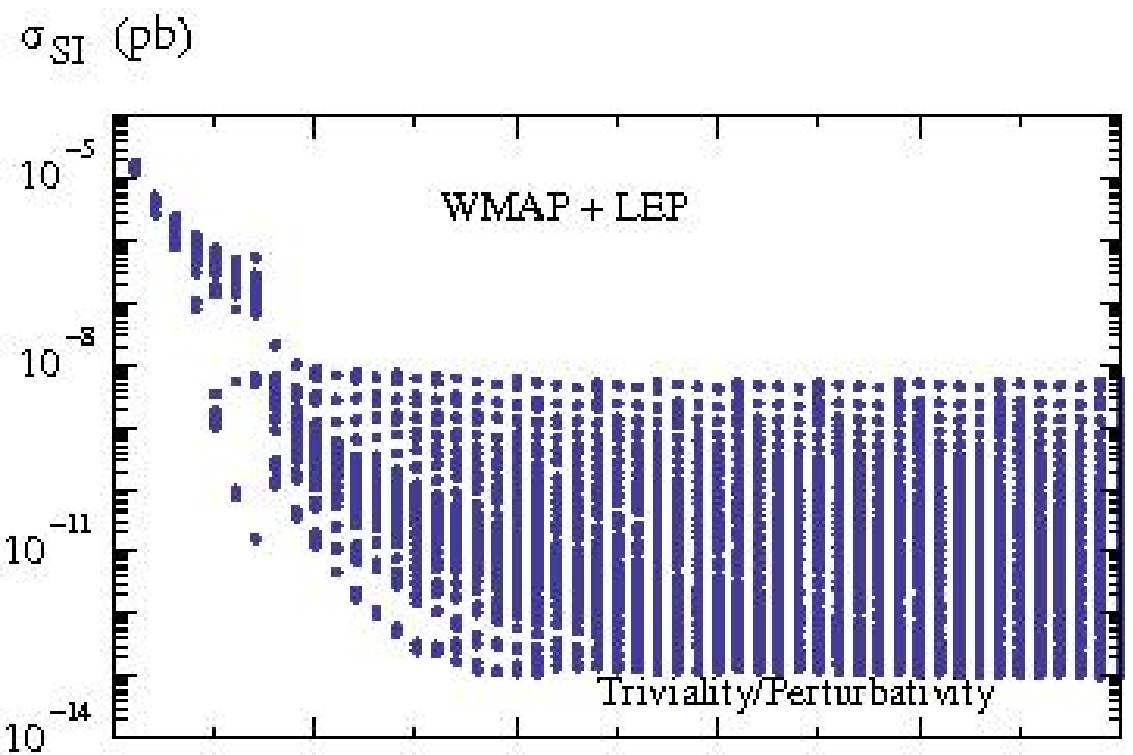}
      \includegraphics[width=3.5in]{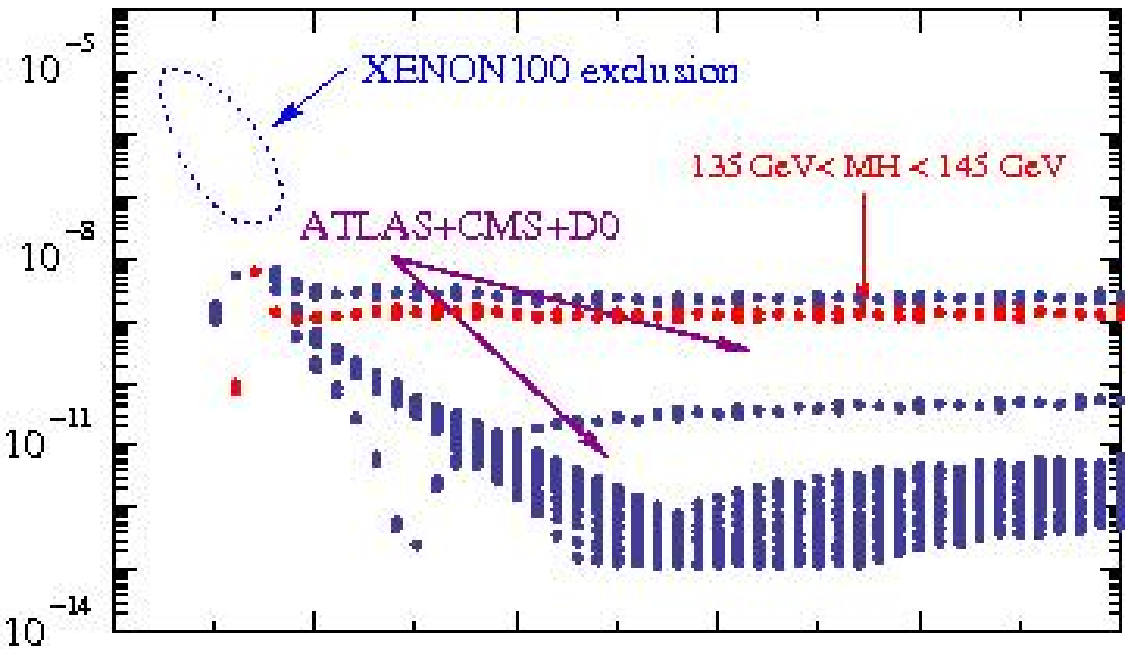}
         \includegraphics[width=3.63in]{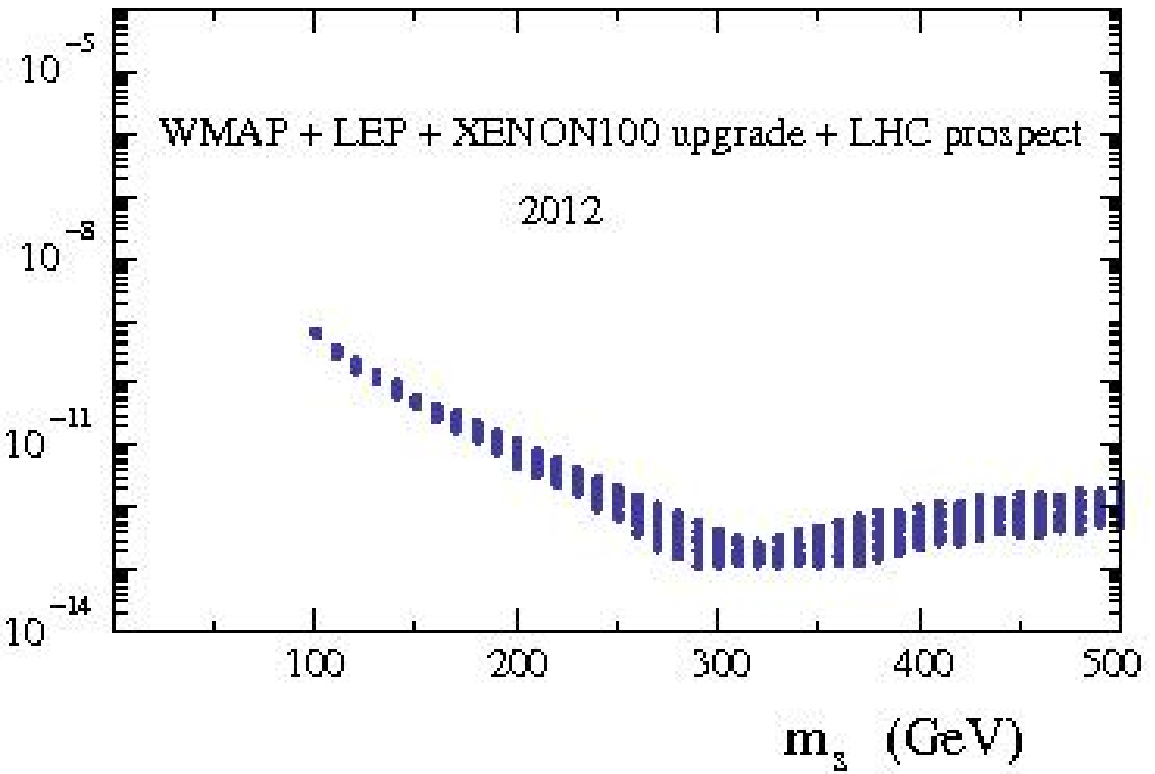}

          \caption{{\footnotesize
Parameter space in the plane ($m_s;\sigma^{SI}_{s-p}$) still allowed after combining different constraints : 
WMAP, LEP and triviality/perturbativity bound (top);
 XENON100, CMS, ATLAS and D0 (middle);
and prospect from ATLAS and XENON100 upgraded in the 5-30 keV range for 2012 (bottom). 
We also show in red the region favored for a would-be Higgs
boson mass $135 \mrm{~GeV} \lesssim M_H \lesssim 155 \mrm{~GeV}$.
}}
\label{Fig:ScanFin}
\end{center}
\end{figure}

 The result of the combination of the set of constraints we just discussed is presented in Fig.\ref{Fig:ScanFin}.
 We see that the influence of LEP constraint on the Higgs mass (114 GeV) excludes a large part of the parameter
  space above $\sigma_{S-p}^{SI}\gtrsim 10^{-44}~\mrm{cm^{2}}$ whereas the triviality/perturbativity bound
  forbids $\sigma_{S-p}^{SI} \lesssim 10^{-49} ~\mrm{cm^2}$ (Fig.\ref{Fig:ScanFin} top). In the meanwhile, 
  the XENON100 data exclude
  the region of low DM mass and high spin--independent cross section (Fig. \ref{Fig:ScanFin} middle).
   Once one includes the CMS/ATLAS/D0 analysis,
  two large holes appear in the parameter space, which will be reduced in one hole by the end of the year
  with the CMS/ATLAS combined analysis. We plotted with red dots the parameter space corresponding
  to a Higgs boson mass $135 < M_H < 155$ GeV (see the following section for more details).
  
  \noindent
  We also show in Fig.\ref{Fig:ScanFin} (bottom)  the prediction expected for the next year (2012) taking
   into account the projected sensitivity 
 of XENON100 experiment  \cite{Aprile:2009yh} in an upgraded version of the detector where 
 the PMT array will be replaced
 by quartz photon intensifying detectors. Its sensitivity along with the projected 1 ton sensitivity are 
 presented in Fig.\ref{Fig:Luminosity} (right).
 We also took into account the projections for the ATLAS sensitivity to the Standard Model Higgs 
 boson from LHC running at centre-of-mass energies of $\sqrt{s}$ = 7 TeV \cite{LumiATLAS}.
 This study extends the previous results of the collaboration by considering the luminosities 
 required to reach 5$\sigma$ discovery significance.
 The result of the analysis  is presented in Fig.\ref{Fig:Luminosity} (left) considering a $5\sigma$ Higgs discovery.
 We had to normalize this sensitivity ${\cal L} \rightarrow {\cal L}/(1-\mrm{Br^{inv}_H})^2$ to take into account
 the invisible width of the Higgs\footnote{Which is negligible in the large part of the parameter space allowed by XENON100.}.
 Once we take into consideration the whole set of predicted sensitivities
 we observe (Fig.\ref{Fig:ScanFin} bottom) that the main part of the parameter space should be easily covered by the end of next year.
 The only region which can escape the observation would be the one corresponding to the limit of perturbativity.
  We should then wait the shutdown and upgraded version of the LHC to cover the $entire$ parameter space of the model.
 
 One of the keypoints of this analysis is based on the fact that the LHC collaborations will be able to reach a part of the parameter 
 space which would never be reached by direct detection technologies 
 because in the region of heavy higgs  ($M_H \gtrsim 300$ GeV) one expect a very low direct detection rate
 ( $\sigma^{SI}_{S-p} \lesssim 10^{-47} \mrm{cm^2}$).
In the meantime, the XENON100 experiment
 can exclude a part of the parameter space ($M_H \lesssim 118$ GeV) which would necessitate a very high luminosity
 to be observed by ATLAS or CMS (see Fig.\ref{Fig:Luminosity}).

\section{Higgs signal?}

Recently the ATLAS collaboration quantified a $2.5 \sigma$ excess for a Higgs boson 
mass $140-150$ GeV. This is coherent when combined with CMS and D0 result, corresponding
to a mean value $M_H \simeq 145$ GeV.
We plotted this "discovery" parameter space in Fig.\ref{Fig:ScanFin}.
The points in red respect  WMAP and XENON100 constraints, in the range 
135 GeV $< M_H <$ 145 GeV\footnote{A similar analysis restricted to a region of parameter space where invisible decay width
of the Higgs
reaches 40 percents has been developed in \cite{Raidal:2011xk}}.
 From this region of parameter space, and including all
the previous reliable constraints, one can deduce that 
$8 \times 10^{-46} \mrm{cm^2}  \lesssim \sigma_{s-p}^{SI}\lesssim 2 \times 10^{-45} \mrm{cm^{2}}$. 
Having a look at the Fig.\ref{Fig:Luminosity}, we can see that this region will unluckily not be reached by an upgraded version
 of the XENON100 experiment next year. In the meantime, a 1 ton extension would easily cover this region
 of the parameter space and will probe the singlet scalar dark matter paradigm, except in a very small region
 of the parameter space, where $m_S \lesssim 100$ GeV which will be very difficult to observe with a XENON--like
 experiment.

\noindent
However, one of the main issue is that the scattering cross section is independent of the DM mass.
Indeed, if we combine Eq.(\ref{Eq:SigV}) and Eq.(\ref{Eq:SigSI}) one understands easily that
for a given value of $M_H$ and $\langle \sigma v \rangle$ (and so, to a relatively good approximation,
of $\Omega_S h^2$) $\sigma_{S-p}^{SI}$ is fixed independently of $m_S$ 
once $m_S \gtrsim M_H$. This means that it
will be difficult to determine the scalar mass even in projected direct detection experiments, like a 1T
XENON--like.

\noindent
It is also interesting to point out that the Higgs-portal construction is similar by several 
aspects to the Z'-portal model of dark matter \cite{Z'}:  as any Higgs searches restrict severely
the parameter space of the model, any Z' searches at LHC should be use in complementarity
with direct detection searches to probe the entire parameter space allowed by WMAP.
At the same time, the analysis should be done is SUSY scenario where light Higgses
are the main annihilation channel, leading to sever direct detection constraints \cite{SUSY}.

\noindent
Writing the conclusion of this work, we noticed that authors just looked at some
consequences of recent Higgs searches at LHC in the NMSSM case \cite{Ellwanger:2011sk} 
and extended scalar sectors \cite{He:2011ti}.

\section{Conclusion and prospect}

In this work, we studied the strong complementarity  between the measurement
of elastic scattering of dark matter on nucleon and the Higgs searches at LHC.
We first studied in detail the influence of the new analysis of the strange quark
content of the nucleon, especially from recent lattice results.
We then showed that in a framework where the Standard Model is extended by a singlet scalar dark matter,
combining the last XENON100 experiment data with WMAP saves only the parameter space
where the invisible decay branching ratio of the Higgs $BR^{inv}(H \rightarrow SS) \lesssim 10 \%$ rendering
the Higgs a Standard Model one,except in a small region with very light dark matter 
($\lesssim 10$ GeV) not yet excluded by XENON100 or around 60 GeV
where the ratio can reach 50\% to 60\%. 
 We have then applied the very recent searches of Higgs released by
ATLAS, CMS and D0 and excluded a huge part of the parameter space, which will be tested at 95 \%
by the end of 2012. LHC collaborations will reach a region
 which could never be accessible by any kind of dark matter direct detection orientated experiments. 
Moreover, if one takes seriously the possibility of a hint around $M_H \simeq 145$ GeV, this would imply
a scattering cross section of $\sigma^{SI}_{S-p}\simeq 10^{-45} \mrm{cm^2}$, testable in future upgraded
version of XENON100. In any scenario, the next months of data/analysis will give precious answers to 
all these interrogations.


\section*{Acknowledgements}
The author want to thank particularly J.B Devivie,  M. Tytgat, A. Falkowski, B. Zaldivar
and the Magic Monday Journal Club
 for (very) useful discussions. The work was
supported by the french ANR TAPDMS {\bf ANR-09-JCJC-0146} 
and the spanish MICINNÕs Consolider-Ingenio 2010 Programme 
under grant  Multi- Dark {\bf CSD2009-00064}.

\section*{Appendix : useful formulae.}

\be
\langle \sigma_{f \bar f} v \rangle = \frac{\lambda_{HS}^2 (m_S^2-m_f^2)^{3/2} m_f^2}
{16 \pi m_S^3 [(4m_s^2-M_H^2)^2+M_H^2\Gamma_H^2]}
\label{Eq:SigV}
\ee

\be
\Gamma_H(H \rightarrow SS) = \frac{\lambda_{HS}^2 M_W^2}{32 \pi g^2 M_H^2} \sqrt{M_H^2 - 4 m_S^2}
\label{Eq:GammaH}
\ee

\be
\sigma^{SI}_{S-p} = \frac{m_p^4 \lambda_{HS}^2 (\sum_{q} f_q )^2}{16 \pi (m_p + m_S)^2 M_H^4}
\label{Eq:SigSI}
\ee


\end{document}